# Controlled Nucleation and Growth of Carbon Nitride Films on CNT Fiber Fabric for Photoelectrochemical Applications

Neeta Karjule, Moumita Rana, Menny Shalom, Jesús Barrio* and Juan José Vilatela*

Dr. Neeta Karjule, Prof. Menny Shalom, Dr. Jesús Barrio
Department of Chemistry and Ilse Katz Institute for Nanoscale Science and Technology, Ben-Gurion University of the Negev, Beer-Sheva 8410501, Israel.

Dr. Moumita Rana, Dr. Juan José Vilatela
IMDEA Materials, Getafe, Madrid E-28906, Spain.
E-mail: juanjose.vilatela@imdea.org

Dr. Jesús Barrio
Faculty of Engineering, Department of Materials, Imperial College London, London, UK.
E-mail: j.barrio-hermida@imperial.ac.uk

**Abstract**

The controlled growth of carbon nitride (CN) films with tailored electronic properties and surface area is quite challenging due to the solid-state reaction and the lack of an efficient interaction between C-N monomers and substrates. Herein we report on the controlled growth of CN films over robust carbon nanotube fiber fabric, which is obtained by either direct calcination of melamine on their surface, that yields a bulk material, or by its chemical vapor deposition resulting in hybrid films. These embodiments are effective electrodes consisting of high-surface area CN containing a carbon nanotube fiber fabrics acting as scaffold and highly conducting built-in current collector. The obtained results confirm that carbon nanotubes act as nucleation center for the formation of CN films, forming close contact at the CN/CNT interphase, and resulting in efficient change transfer upon illumination and enhanced electrochemical surface area.

# 1. Introduction

Graphitic carbon nitride (CN) materials have attracted widespread attention during the last decade due to its mechanical robustness, low price and semiconducting properties, which make them suitable materials for environmental and energy-related scenarios such as water splitting,[1,2] $CO_2$ reduction,[3,4] water cleaning,[5] optoelectronic devices[6,7], sensing[8,9] and organic transformations.[10–12] Particularly, CN-based self-standing systems such as films or hydrogels have been recently a focus of research, given the potential capability of being utilized as a substrate for photoelectrocatalytic scenarios with controlled reaction sites and a better recyclability and reutilization.[13–15] Additionally, the combination of carbon nanostructures with CN materials has been proven to yield beneficial synergistic effects arising from the creation of metal-free heterojunctions.[16] Carbon materials such like graphene, fullerenes, or carbon nanotubes (CNT) act as an electron sink improving the charge separation efficiency of the system and the overall photoelectrocatalytic performance.[17] Particularly, macroscopic fibers, yarns or fabrics of carbon nanotubes are attractive conductive scaffolds because of their network structure of porous, yet bundled conducting nanocarbons giving rise to high porosity and high electrical conductivity. Carbon nanotube (CNT) fibers inherently have high surface area (~250 $m^2$ $g^{-1}$), electrical conductivity in the range of metals (on a mass basis), and extraordinary mechanical toughness.[18,19] They can be used directly as electrodes, for example in structural supercapacitors, or as counter electrodes in photoelectrochemical cells.[20–23] In general, growing a nanostructured active phase directly on the internal pores of CNT fibers maximizes surface area of the active phase, while also minimizing charge diffusion length through the active phase, such hybrid structure has proven effective to increase efficiency of multiple (photo)catalytic and electrochemical processes.[24–27]

Hence, in this work we utilized high performance CNT fiber fabric as a substrate for the controlled growth of CN materials, the different synthetic pathway, namely direct calcination of melamine over CNT, or its chemical vapor deposition, allows the preparation of hybrids with different physical properties and applicability. The nucleated growth of CN films over CNT fibers by chemical vapor deposition (CVD) was confirmed by transmission electron microscopy (TEM), and improved charge separation was proven by photophysical characterization as well as photocurrent generation. Furthermore, as a proof of concept, we show the photocatalytic activity of the hybrid in both the powder form as well as the self-standing film.

## 2. Results and discussion

Carbon nanotube fiber (CNT)–carbon nitride (CN) hybrids were prepared by the thermal condensation of melamine directly over CNT fiber at 500 °C under $N_2$ atmosphere for 1h (Scheme S1). The same procedure was carried out utilizing the cyanuric acid – melamine supramolecular hydrogen bonded framework,[28,29] but no CN was deposited over the CNT due to the quench in the sublimation upon condensation as a result of the hydrogen bonds within the structure (Figure S1). The initial characterization of the materials was carried out after mechanically grinding the initial films to powder. Fourier transformed infrared spectroscopy (FTIR) and X-ray diffraction (XRD) analysis confirmed the formation of polymeric CN on the surface of the CNT; namely the stretching vibrations corresponding to the heptazine units are visible within 1700 – 1300 cm$^{-1}$, as well as the breathing mode of triazine rings at 815 cm$^{-1}$ (Figure S2a). The XRD patterns show the diffraction peaks corresponding to in plane repetition unit of the heptazine moieties, the (100) crystal plane, at 13.0° and the inter layer stacking peak, (002) crystal plane at ~27° (Figure S2b). We could observe that the presence of the CNT fiber modifies the *d* spacing (Å) between graphitic CN layers, as the $2\theta$ value is shifted from 27.3 (3.26 Å) to 26.9° (3.31 Å) implying a longer distance between layers (Figure S2b). The chemical states of C and N within CNT-CN were analyzed through X-ray photoelectron spectroscopy (XPS), which reflected all the binding energies corresponding to the CN polymers, along with an enhanced contribution for the C-C chemical bond due to the presence of CNT fiber (Further discussion in Figure S3). The intimate contact between the CN and the CNT fiber was confirmed scanning electron microscopy (SEM), where we could observe that CNT fibers are templating the growth of CN into a tube shape material, and both counterparts are homogeneously distributed along the hybrid (**Figure 1a-b**, Figure S4).[30] Furthermore, cross-section images show the formation of a thick (*ca.* 96 µm) layer of CN (Figure S5). Additionally, atomic force microscopy (AFM) further confirm the interaction between CNT fiber and CN (**Figure 1c-d**).

The interaction between the CNT and CN was reflected in the photophysical properties of the hybrid as well; a comparison of the UV-vis spectra of the hybrid with pristine CN exhibits a prominent enhancement in the optical absorption (Figure S6b). Additionally, fluorescence spectroscopy confirms the improvement of the electron – hole separation due to the formation of a heterojunction between CNT and CN, as observed in the reduced fluorescence intensity (Figure S6a).[31] Those properties make the CNT-CN hybrid material very suitable as a photocatalyst for processes such as the degradation of an organic dye, which we evaluated utilizing Rhodamine B as a model for organic pollutant in an aqueous environment.[32] We confirmed the enhanced photocatalytic activity of the hybrid, as well as the photodegradation mechanism (Further details in figure S7, S8).[33] Nevertheless, the obtained CN coating on CNT fibers is very thick and does not allow an easy manipulation as self-standing films, therefore, for the expansion of the CNT-CN hybrids in other energy-related scenarios, and in order to obtain a self-standing material with a better processability we carried out the synthesis of CN films over CNT fibers by chemical vapor deposition (CVD) of melamine (**Figure 2a**).[34–36]

The chemical states of C and N in the hybrid films were analyzed by depth profile XPS measurements (Figure 2b-c). We utilized Ar etching at a speed of 0.07 nm s$^{-1}$ in order to access the inner CN layers in contact with the CNT fiber. The chemical states of the elements in the surface resemble the typical contributions for the pristine CN materials, namely adventitious C-C bonds at 284.8 eV, C-N=C coordination at 288.3 eV, and contributions corresponding to oxygen absorbed in the surface and other impurities.[37] Upon sputtering up to 17.5 nm depth, we could observe a general shift to lower binding energies due to a strong interaction of the CN with the CNT fiber (Table S1). Additionally, the ratio between the C-N=C and C-C contribution decreases from 6.2 to 2.8 indicating a higher proportion of CNT in the inner levels. In the same manner, the N1s spectra, which displays three different chemical binding energies corresponding to C-N=C, N-C$_3$, and N-H groups, shows a shift of the contributions to lower values in the deeper levels owing to the interaction with the CNT fiber. SEM images taken at the edge between CNT fiber and vapor-growth CN shows a gradient of coverage of individual CNT fibers by CN (**Figure 3**), confirming that polymeric CN is growing on the CNT substrate. SEM images taken in the middle of the CNT-CN films reveal the formation of ordered CN architectures of several micrometer length growing perpendicular to the CNT fiber (Figure S9). Cross-section SEM images show a uniform and thick CN layer (*ca.* 77 μm) on the CNT fiber (Figure S10).

To understand the microstructure of the CNT-CN hybrid in further details, we performed TEM analysis (**Figure 4**). The samples analyzed were extracted from the middle of the 5-micron thick sample in order to avoid a possible overlayer of amorphous CN, as typically formed in the CVD processes upon precursor decomposition during cooling. As shown in Figure 4 the hybridization of the fibers with CN resulted in a uniform coating around the CNT bundles throughout their length. Figure 4c and d show the high resolution TEM images of the hybrid. The carbon nitride coating is a combination of amorphous and crystalline regions. The latter are readily visible through the (002) planes of the carbon nitride. Interestingly, the crystalline regions of CN near CNT bundles are preferentially oriented parallel to the CNT growth direction, indicating that CN growth is probably assisted by nucleation on the graphitic walls of the CNTs. It is unclear if there is epitaxial match between the two phases, but accelerated nucleation on CNT fibers is generally observed even in the absence of epitaxial growth.[38] Very importantly, the observed orientational arrangement implies that in the event of interfacial charge transfer, it must occur through overlapping π orbitals of the CN sheets of the CNT walls.

We performed Raman spectroscopy of the CNT-CN hybrid to gain further insight into the coupling of CNT and CN (**Figure 5** and S11). As shown in the Figure 5a, three main peaks at 1319.7 (±2.3), 1584.3 (±0.9) and 2606 (±2.6) cm$^{-1}$ were observed. The broad peak around 1320 cm$^{-1}$ can be deconvoluted to three smaller peaks. Among them the peak at 1318.3 cm$^{-1}$ is originated from the vibrations from the defected graphitic network of the carbon nanotubes (D band). The smaller peaks at 1244.8 and 1400 cm$^{-1}$ can be related to the symmetric stretching of the C-N and C=N bonds.[39–41] The strong peak at 1584.3 cm$^{-1}$ can be a combination of the symmetric graphitic vibrations from the CNT fibers as well as CN. The small, isolated peak at 2606 cm$^{-1}$ corresponds to the 2D band of the CNTs. Features from both CNT and CN were observed in all spectra, confirming uniform distribution of the two phases in the hybrid. The ratio of D/G intensity in the CNT-CN hybrid increased to 0.3 from a value of 0.13 in pristine CNT fibers, on account of a more disordered structure of the newly deposited carbon nitride in the hybrid. Raman measurements were also taken for reference on a CN overlayer, which exhibited its typical features,[42,43] located

in the range of 1000-1600 cm$^{-1}$ (1086, 1162, 1243,1322, 1410, 1482 and 1581 cm$^{-1}$), significantly differing from those found at the pristine CNT.

A particular aspect of interest is to determine a possible strain and/or charge transfer in the CNT/CN hybrids, as do occur in CNT/inorganic hybrids.[44] We observe no shift in the G band (Figure 5b), implying that there is no net charge transfer nor strain induced upon hybridization of CNTs with CN.[45,46] Since the crystalline CN domains are non-covalently attached and have close crystallographic match with graphite, it is not surprising that hybridization does not induce deformation of the stiff CNTs.

On the other hand, while the Raman spectra shows not net charge transfer, it does not rule out local charge transfer from CN layers, particularly their amine, pyridinic and quaternary nitrogen centers, which are known to produce charge redistribution in graphitic materials.[47,48] The presence of local electron-rich and hole-rich regions in a continuous carbon matrix is considered to favor separation of the photoexcited electrons and holes, thus inhibiting their recombination and increasing efficiency of some photocatalytic processes.[49] Such mechanism does not rely on a strong CNT-CN interaction involving charge transfer upon hybridization. Literature examples of combinations of CN and nanocarbons with demonstrated improvements in photocatalytic activity in general show no appreciable change in the Raman G band.[50,51] Interestingly, the CNT-CN hybrids in this work do present an appreciable downshift of the 2D band of around 5 cm$^{-1}$. Without a corresponding downshift change in the G band it cannot be taken as indication of doping or strain. Speculatively, this feature may arise from interaction of the CNT with CN through local electron-rich and hole-rich regions.

As a proof of concept we utilized the CNT-CN self-standing films as a photoanode in a standard water splitting PEC cell.[52,53] The CNT-CN fibers were attached to FTO substrate (unless specified otherwise) with a Cu wire. We could observe that upon illumination the films showed a stable photocurrent of 8.5 µA (at 1.23 V vs RHE) with certain dark current due to the intrinsic capacitance of the materials, while bare CNT had no activity (**Figure 6a**, S12). In order to evaluate purely the contributions belonging to the CNT-CN electrode and discarding the interferences coming from FTO, we performed the same measurement attaching the CNT-CN electrode to bare glass instead to FTO and observed that the measured photocurrent displayed a significant increase up to 14.5 µA (at 1.23 V vs RHE) with a consequent enhancement of the dark current (Figure S13). Triethanolamine (TEOA), a widely utilized hole scavenger was added to the electrolyte for ensuring a maximum hole extraction. In this scenario, the photocurrent was enhanced up to 17.2 µA, which implies and increase in charge transfer efficiency of almost 50% for CNT-CN (in FTO). Additionally, the stability was improved in the presence of TEOA, showing less fluctuation during nearly 2-hour measurement (Figure S14). Furthermore, the photoanodic performance of the CNT-CN anodes was evaluated by linear sweep voltammetry measurements, where we observed that upon illumination CNT-CN displays a cathodic shift in the onset potential which is enhanced upon addition of TEOA in the electrolyte, confirming the charge transfer and separation efficiency of the CNT-CN films (Figure 6b).

In order to gain deeper insights into the interfacial charge transfer resistance ($R_{ct}$) of the materials, we conducted electrochemical impedance spectroscopy (EIS) studies (Figure S15a and Table S2).[54] The equivalent circuit used to fit the Nyquist plots is depicted in Figure S15b. It is observed that the EIS radius is smaller for CNT-CN than for CN-M (electrode prepared using FTO as substrate). In the case of CNT-CN, the diameter of the arc becomes smaller, suggesting lower $R_{ct}$ which confirms the efficient separation and migration of photoinduced charge carriers.

The electrochemical surface area (ECSA) of the materials is proportional to the electrochemical double-layer capacitance ($C_{dl}$), which can be determined by cyclic voltammetry at different scan rates (Figure 15c and Table S2).[55] The measured $C_{dl}$ are 0.027 and 1.61 mF cm$^{-2}$ for CN-M and CNT-CN, respectively. Therefore, the ECSA is 60 times higher when utilizing CNT fiber as substrate instead of FTO. This indicates that additional electrochemical active sites are created, which also contributes to the improved PEC performance. These results suggest that the CNT-CN exhibits faster electron transfer kinetics, larger electroactive surface area and better electronic conductivity than CN-M. In accordance with ECSA, the N$_2$ sorption analysis (Figure S16) showed a slight enhancement of the specific surface area for CNT-CN films (grown by CVD of melamine, $S_A = 9.93$ m$^2$ g$^{-1}$) relative to the CN powder counterpart (CN-M, $S_A = 6.75$ m$^2$ g$^{-1}$).

## 3. Conclusion

In summary, in this work we have prepared CNT-CN powder hybrid and self-standing films by direct calcination or chemical vapor deposition of melamine on CNT fiber fabric, respectively. Photophysical characterization as well as TEM and Raman confirm the strong interaction between both counterparts leading to vertical growth and charge transfer upon illumination. The self-standing electrodes are utilized in a standard water splitting photo-electrochemical cell as proof of concept showing stable generation of photocurrent and a substantial enhancement of the electrochemical surface area due to the CNT-templated growth of CN versus the CN prepared in FTO glass. We believe that this work opens the gate towards further research in metal-free self-standing photoelectrochemical systems as well as for the utilization of high-performance CNT fiber as a template and nucleation center for the growth of diverse functional materials.

## 4. Experimental Section

**4.1. Synthesis of CNT fiber.** The CNT fibers were synthesized by a floating catalyst chemical vapor deposition process using toluene, ferrocene and thiophene as carbon, catalysts and promoter sources, respectively. Briefly, these precursors were introduced in a vertical furnace at a temperature of 1250 °C under hydrogen atmosphere. Under these conditions, an aerogel of long CNTs is formed floating in the gas phase. By mechanically withdrawing it from the gas phase during constant precursor injection, continuous macroscopic fibers of CNTs are formed. Winding multiple filaments on a rotating spool leads to a unidirectional non-woven fabric, essentially a porous, conductive electrode like a sheet of paper. The CNT fiber material used in this work was in such fabric format.

**4.2. Synthesis of CN and CNT-CN materials.** Melamine powder was directly placed homogeneously on top of CNT fiber fabric in a ceramic crucible. The crucible was heated up to 500 °C for 1 h under $N_2$ flow with a heating ramp of 2.3 °C $min^{-1}$. In order to create CNT-CN self-standing electrodes, melamine was placed in the bottom of a ceramic crucible with CNT fiber on top of it, covered with aluminum foil. Upon thermal treatment at 500 °C for 1 h with a heating ramp of 2.3 °C $min^{-1}$ under $N_2$ flow, polymeric CN is deposited in the CNT fiber. CN reference powder was prepared by heating melamine in a ceramic crucible for 4 h at 500 °C under $N_2$ atmosphere. Reference CN films on FTO were prepared by doctor blading a paste of melamine and ethylene glycol over clean FTO glass and subsequent thermal condensation at 550 °C for 4 h under $N_2$ atmosphere as reported previously.[56]

**4.3. Photocatalytic Rhodamine B degradation.** The photocatalytic activity of the material was assessed by performing the rhodamine B (RhB) dye degradation under a white 100 W LED (Bridgelux BXRA-50C5300; $\lambda >$ 410 NM) illumination. The prepared materials (20 mg) were mixed along with 20 mL of RhB solution (20 mg L$^{-1}$) in water in the dark until an adsorption-desorption equilibrium was stablished. The reaction starts after switching on the LED and the RhB remaining concentration is monitored by measuring the optical absorbance at its $\lambda_{max}$ (554 nm).

**4.4. Photoelectrochemical (PEC) measurements.** The PEC properties of as-prepared CN electrodes were measured in a standard PEC cell. All electrochemical measurements were acquired using a three-electrode system on an Autolab potentiostat (Metrohm, PGSTAT 101). A Pt-foil counter electrode and Ag/AgCl (in saturated KCl) reference electrode were utilized. The electrolyte was a 0.1 M KOH aqueous solution (pH ~13), and for analyzing the performance with 100% hole extraction, a 0.1 M KOH aqueous solution with 10% (v/v) triethanolamine (TEOA) was utilized. The potentials for chronoamperometric techniques were converted to the reversible hydrogen electrode (RHE). Photocurrent densities were measured at 1.23 V vs RHE under the irradiation of a solar simulator at a power density of 100 mW cm$^{-2}$ (Newport 300 W Xe arc lamp, equipped with air mass AM 1.5 G and water filters). Nyquist plots of the samples were measured in the frequency range from 100 kHz to 10 mHz at an applied voltage of 1.23 V vs RHE.

**4.5. Characterization.** SEM images were obtained with a JEOL JSM-7400, FEG source operating at 3.5 kV. Fourier-transformed infrared spectroscopy (FTIR) was performed on a Thermo Scientific Nicolet iS5 FTIR spectrometer (equipped with a Si ATR). Transmission electron microscopy (TEM) images were recorded using a Talos F200X FEG microscope, operated at 80 kV. Fluorescence measurements were carried out on FluroMax®4 spectrofluorometer from Horiba Scientific. UV-vis absorption spectra were obtained with a Cary 100 spectrophotometer. X-Ray photoelectron spectroscopy (XPS) was performed with a Thermo Fisher Scientific ESCALAB 250 using monochromatic K$\alpha$ X-rays (1456.6 eV). The depth profile of the sample was obtained by combining a sequence of Ar ion gun etch cycles interleaved with XPS measurements from the current surface. The sputtering rate was approximately 0.07 nm·s$^{-1}$. The XPS results were processed by using the AVANTGE software. X-ray diffraction patterns (XRD) were obtained using a PANalytical's Empyrean diffractometer equipped with a position sensitive detector X'Celerator. Data was collected with a scanning time of ~7 min for 2$\theta$ ranging from 5° to 60° using Cu K$\alpha$ radiation ($\lambda$= 1.54178 Å, 40 kV, 30 mA). Raman measurements were performed using Ranishaw instrument, fitted with a 785 nm laser source. Measurements were taken on samples peeled with tape in order to remove a possible over-layer of CN formed on the top surface of the hybrid. Nitrogen-sorption measurements and Brunauer-Emmet-Teller (BET) specific surface area calculations were performed using Quantachrome NOVAtouch NT LX$^3$ system.

**Supporting Information**

Supporting Information is available from the Wiley Online Library or from the author.

**Acknowledgments**

This research was funded by the Israel Science Foundation (ISF), grant No. 1161/17, and supported by the Minerva Center No. 117873. We thank Mr. Juergen Jopp from the scientific staff of the Ilse Katz Institute for Nanoscale Science and Technology for assistance with AFM measurements, and to Dr. Jiawei Xia and Ms Liel Abisdris for assistance in materials characterization. Financial support is acknowledged from the European Union Seventh Framework Program under grant agreement 678565 (ERC-STEM), the Clean Sky Joint Undertaking 2, Horizon 2020 under Grant Agreement Number 738085 (SORCERER), from MINECO (RyC-2014-15115, Spain) and HYNANOSC (RTI2018-099504-A-C22), as well the FOTOART-CM project funded by the Madrid Region under program P2018/NMT-4367.

## References

[1] X. Wang, K. Maeda, A. Thomas, K. Takanabe, G. Xin, J. M. Carlsson, K. Domen, M. Antonietti, *Nat. Mater.* **2009**, *8*, 76.
[2] F. Wu, Y. Ma, Y. Hang Hu, *ACS Appl. Energy Mater.* **2020**, *3*, 11223.
[3] J. Barrio, D. Mateo, J. Albero, H. García, M. Shalom, *Adv. Energy Mater.* **2019**, *9*, 1902738.
[4] S. N. Talapaneni, G. Singh, I. Y. Kim, K. AlBahily, A. H. Al-Muhtaseb, A. S. Karakoti, E. Tavakkoli, A. Vinu, *Adv. Mater.* **2020**, *32*, 1904635.
[5] F. Opoku, K. K. Govender, C. G. C. E. van Sittert, P. P. Govender, *Adv. Sustain. Syst.* **2017**, *1*, 1700006.
[6] J. Xu, M. Shalom, *ChemPhotoChem* **2019**, *3*, 170.
[7] F. K. Kessler, Y. Zheng, D. Schwarz, C. Merschjann, W. Schnick, X. Wang, M. J. Bojdys, *Nat. Rev. Mater.* **2017**, *2*, 17030.
[8] M. M. Xavier, P. R. Nair, S. Mathew, *Analyst* **2019**, *144*, 1475.
[9] J. Barrio, M. Volokh, M. Shalom, *J. Mater. Chem. A* **2020**, *8*, 11075.
[10] S. Shi, Z. Sun, C. Bao, T. Gao, Y. H. Hu, *Int. J. Energy Res.* **2020**, *44*, 2740.
[11] Y. Markushyna, C. A. Smith, A. Savateev, *European J. Org. Chem.* **2020**, 10, 1294-1309.
[12] I. Ghosh, J. Khamrai, A. Savateev, N. Shlapakov, M. Antonietti, B. König, *Science (80-. ).* **2019**, *365*, 360 LP.
[13] J. Sun, B. V. K. J. Schmidt, X. Wang, M. Shalom, *ACS Appl. Mater. Interfaces* **2017**, acsami.6b14879.
[14] C. Hu, Y.-R. Lin, H.-C. Yang, *ChemSusChem* **2019**, *12*, 1794.
[15] Q. Cao, B. Kumru, M. Antonietti, B. V. K. J. Schmidt, *Mater. Horizons* **2020**, *7*, 762.
[16] J. Fu, J. Yu, C. Jiang, B. Cheng, *Adv. Energy Mater.* **2018**, *8*, 1.
[17] Q. Han, N. Chen, J. Zhang, L. Qu, *Mater. Horizons* **2017**, *4*, 832.
[18] E. Senokos, V. Reguero, J. Palma, J. J. Vilatela, R. Marcilla, *Nanoscale* **2016**, *8*, 3620.
[19] J. J. Vilatela, R. Marcilla, *Chem. Mater.* **2015**, *27*, 6901.
[20] A. Monreal-Bernal, J. J. Vilatela, R. D. Costa, *Carbon N. Y.* **2019**, *141*, 488.
[21] M. Rana, Y. Ou, C. Meng, F. Sket, C. González, J. J. Vilatela, *Multifunct. Mater.* **2020**, *3*, 015001.
[22] A. Moya, A. Cherevan, S. Marchesan, P. Gebhardt, M. Prato, D. Eder, J. J. Vilatela, *Appl. Catal. B Environ.* **2015**, *179*, 574.
[23] D. Liu, M. Zhao, Y. Li, Z. Bian, L. Zhang, Y. Shang, X. Xia, S. Zhang, D. Yun, Z. Liu, A. Cao, C. Huang, *ACS Nano* **2012**, *6*, 11027.
[24] D. H. Kweon, M. S. Okyay, S.-J. Kim, J.-P. Jeon, H.-J. Noh, N. Park, J. Mahmood, J.-B. Baek, *Nat. Commun.* **2020**, *11*, 1278.
[25] M.-Z. Yang, Y.-F. Xu, J.-F. Liao, X.-D. Wang, H.-Y. Chen, D.-B. Kuang, *J. Mater. Chem. A* **2019**, *7*, 5409.
[26] M. Rana, V. Sai Avvaru, N. Boaretto, V. A. de la Peña O'Shea, R. Marcilla, V. Etacheri, J. J. Vilatela, *J. Mater. Chem. A* **2019**, *7*, 26596.

[27]  K. Woan, G. Pyrgiotakis, W. Sigmund, *Adv. Mater.* **2009**, *21*, 2233.
[28]  M. Shalom, S. Inal, C. Fettkenhauer, D. Neher, M. Antonietti, *J. Am. Chem. Soc.* **2013**, *135*, 7118.
[29]  J. Barrio, M. Shalom, *ChemCatChem* **2018**, *10*, 5573.
[30]  X. Chen, H. Wang, R. Meng, M. Chen, *ChemistrySelect* **2019**, *4*, 6123.
[31]  Y. Gong, J. Wang, Z. Wei, P. Zhang, H. Li, Y. Wang, *ChemSusChem* **2014**, *7*, 2303.
[32]  L. Shi, L. Yao, W. Si, *J. Mater. Sci. Mater. Electron.* **2019**, *30*, 1714.
[33]  J. Barrio, N. Karjule, J. Qin, M. Shalom, *ChemCatChem* **2019**, *11*, 6295.
[34]  J. Bian, L. Xi, C. Huang, K. M. Lange, R.-Q. Zhang, M. Shalom, *Adv. Energy Mater.* **2016**, *6*, 1600263.
[35]  P. Giusto, D. Cruz, T. Heil, H. Arazoe, P. Lova, T. Aida, D. Comoretto, M. Patrini, M. Antonietti, *Adv. Mater.* **2020**, *32*, 1908140.
[36]  N. Karjule, J. Barrio, L. Xing, M. Volokh, M. Shalom, *Nano Lett.* **2020**, *20*, 4618.
[37]  T. Y. Ma, S. Dai, M. Jaroniec, S. Z. Qiao, *Angew. Chemie Int. Ed.* **2014**, *53*, 7281.
[38]  H. Yue, A. Monreal-Bernal, J. P. Fernández-Blázquez, J. Llorca, J. J. Vilatela, *Sci. Rep.* **2015**, *5*, 16729.
[39]  E. Quirico, G. Montagnac, V. Lees, P. F. McMillan, C. Szopa, G. Cernogora, J.-N. Rouzaud, P. Simon, J.-M. Bernard, P. Coll, N. Fray, R. D. Minard, F. Raulin, B. Reynard, B. Schmitt, *Icarus* **2008**, *198*, 218.
[40]  Y. H. Cheng, B. K. Tay, S. P. Lau, X. Shi, X. L. Qiao, J. G. Chen, Y. P. Wu, C. S. Xie, *Appl. Phys. A* **2001**, *73*, 341.
[41]  W. Zhang, H. Huang, F. Li, K. Deng, X. Wang, *J. Mater. Chem. A* **2014**, *2*, 19084.
[42]  T. Suter, V. Brázdová, K. McColl, T. S. Miller, H. Nagashima, E. Salvadori, A. Sella, C. A. Howard, C. W.M. Kay, F. Corà, P. F. McMillan, *J. Phys. Chem. C* **2018**, *122*, 25183.
[43]  W.-J. Ong, L.-L. Tan, S.-P. Chai, S.-T. Yong, *Dalt. Trans.* **2015**, *44*, 1249.
[44]  A. Moya, N. Kemnade, M. R. Osorio, A. Cherevan, D. Granados, D. Eder, J. J. Vilatela, *J. Mater. Chem. A* **2017**, *5*, 24695.
[45]  N. S. Mueller, S. Heeg, M. P. Alvarez, P. Kusch, S. Wasserroth, N. Clark, F. Schedin, J. Parthenios, K. Papagelis, C. Galiotis, M. Kalbáč, A. Vijayaraghavan, U. Huebner, R. Gorbachev, O. Frank, S. Reich, *2D Mater.* **2017**, *5*, 15016.
[46]  J. E. Lee, G. Ahn, J. Shim, Y. S. Lee, S. Ryu, *Nat. Commun.* **2012**, *3*, 1024.
[47]  A. Du, S. Sanvito, Z. Li, D. Wang, Y. Jiao, T. Liao, Q. Sun, Y. H. Ng, Z. Zhu, R. Amal, *J. Am. Chem. Soc.* **2012**, *134*, 4393.
[48]  Y. Zhang, T. Mori, L. Niu, J. Ye, *Energy Environ. Sci.* **2011**, *4*, 4517.
[49]  Z. Ma, R. Sa, Q. Li, K. Wu, *Phys. Chem. Chem. Phys.* **2016**, *18*, 1050.
[50]  R. C. Pawar, S. Kang, S. H. Ahn, C. S. Lee, *RSC Adv.* **2015**, *5*, 24281.
[51]  S. K. Lai, C. Xie, K. S. Teng, Y. Li, F. Tan, F. Yan, S. P. Lau, *Adv. Opt. Mater.* **2016**, *4*, 555.
[52]  E. J. Son, S. H. Lee, S. K. Kuk, M. Pesic, D. S. Choi, J. W. Ko, K. Kim, F. Hollmann, C. B. Park, *Adv. Funct. Mater.* **2018**, *28*, 1.
[53]  M. Volokh, G. Peng, J. Barrio, M. Shalom, *Angew. Chemie Int. Ed.* **2019**, *58*, 6138.
[54]  L. Zhang, X. He, X. Xu, C. Liu, Y. Duan, L. Hou, Q. Zhou, C. Ma, X. Yang, R. Liu, F. Yang, L. Cui, C. Xu, Y. Li, *Appl. Catal. B Environ.* **2017**, *203*, 1.
[55]  C. C. L. McCrory, S. Jung, I. M. Ferrer, S. M. Chatman, J. C. Peters, T. F. Jaramillo, *J. Am. Chem. Soc.* **2015**, *137*, 4347.
[56]  G. Peng, L. Xing, J. Barrio, M. Volokh, M. Shalom, *Angew. Chemie Int. Ed.* **2018**, *57*,

1186.

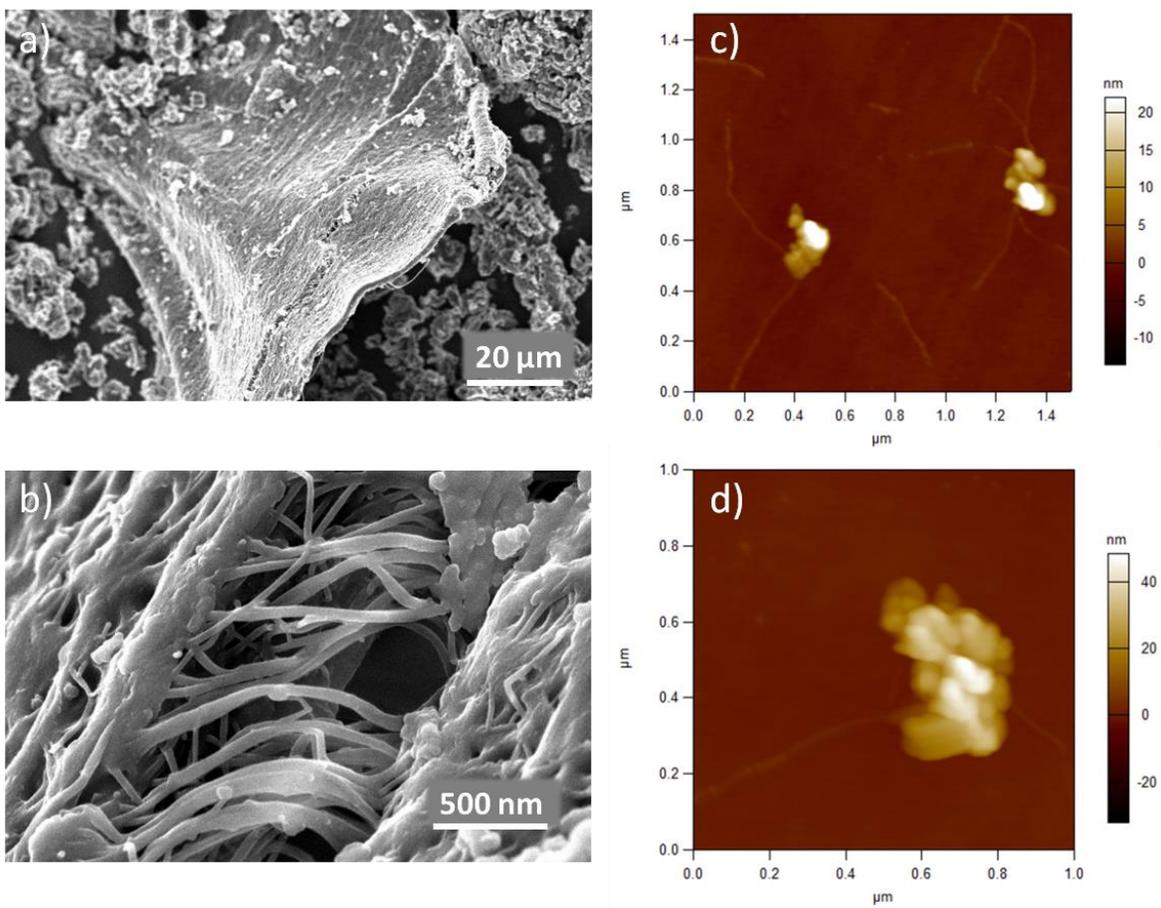

**Figure 1.** Microstructural Characterization of CNT-CN. FESEM images of CNT-CN, showing coating of CN throughout the CNT fibers (a, b) and AFM profiles of CNT-CN (c, d).

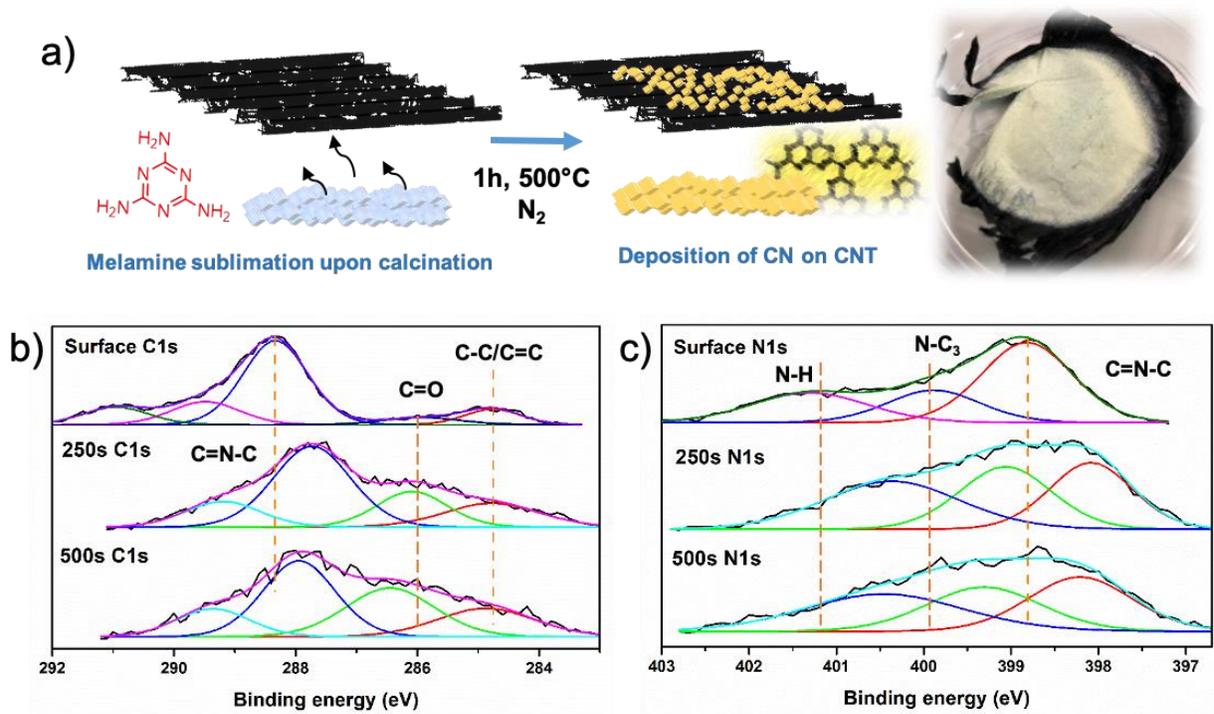

**Figure 2.** Schematic representation of the preparation of CNT-CN films through vapor-solid thermal conversion of melamine (a), C1s depth profile XPS measurements (b), and N1s depth profile XPS measurements (c).

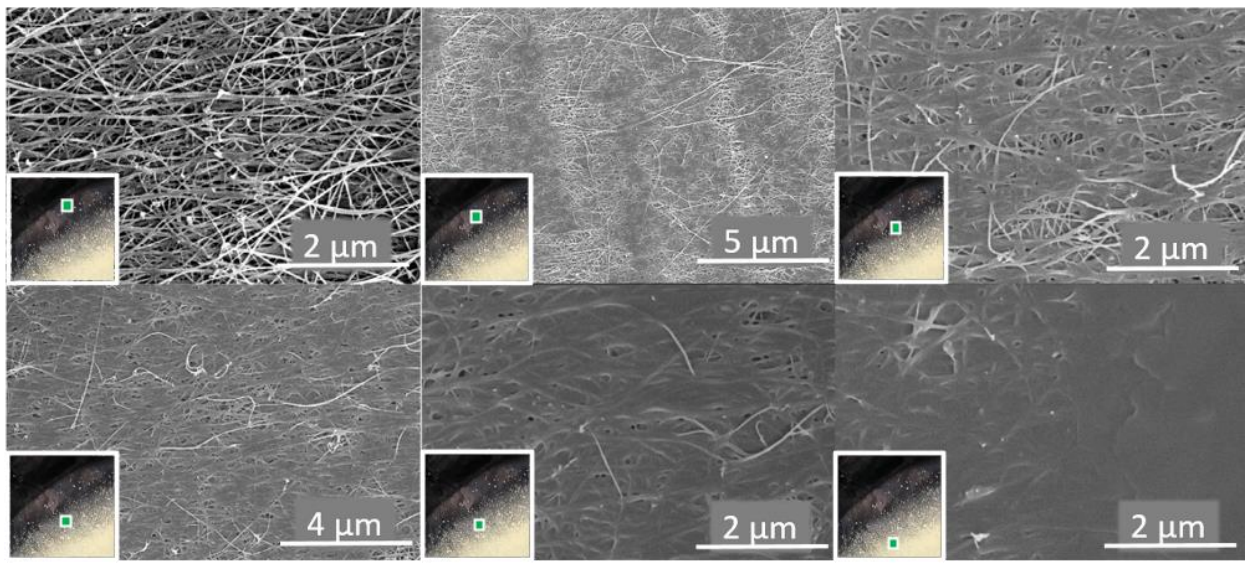

**Figure 3.** SEM images of the CNT-CN fiber in different locations along the film.

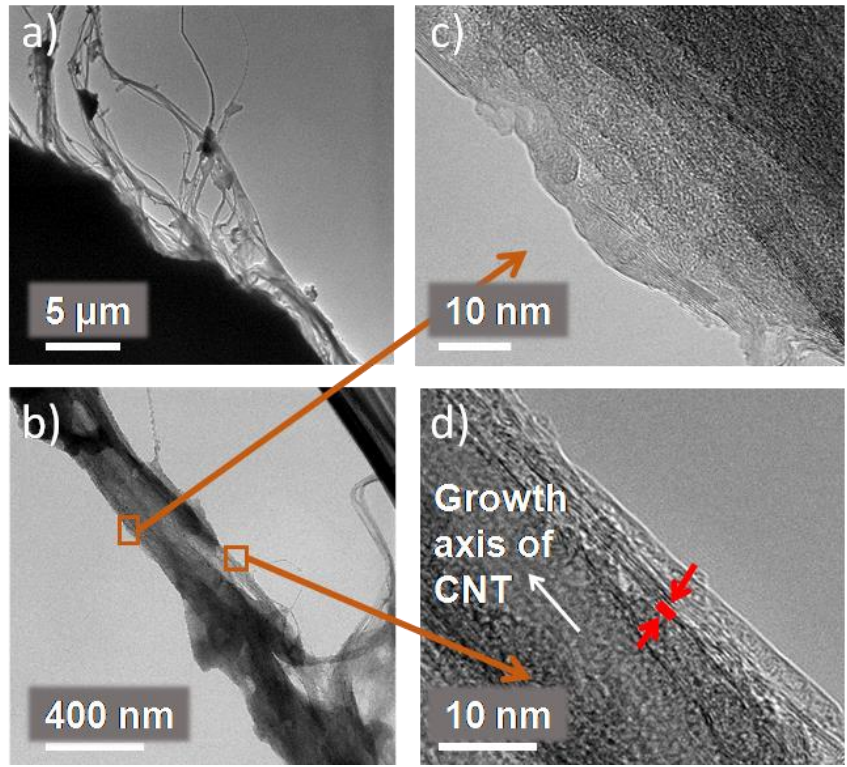

**Figure 4.** Low resolution TEM images of CNT-CN (a, b). High resolution TEM images of CNT-CN (c, d), the marked areas in (d) showing nearly parallel growth of the (002) plane of the CN with the (002) plane/growth axis of the CNT.

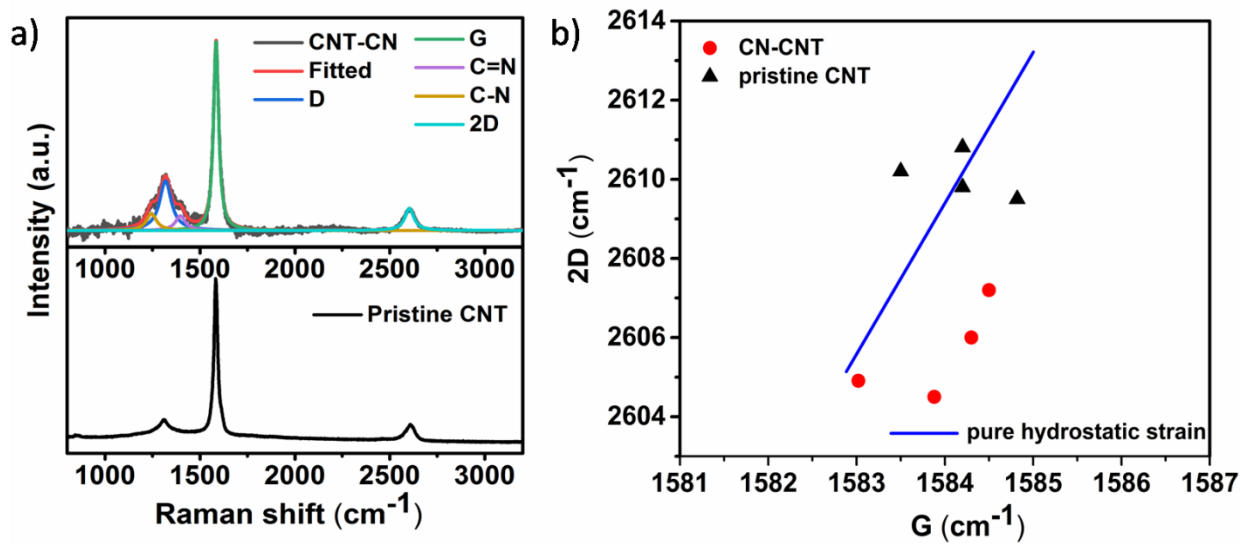

**Figure 5.** Comparison of the Raman spectra of the CNT-CN hybrid with pristine CNT (a). 2D vs G band position showing no induced strain or doping, but an unusual downshift of the 2D band in the hybrid (b).

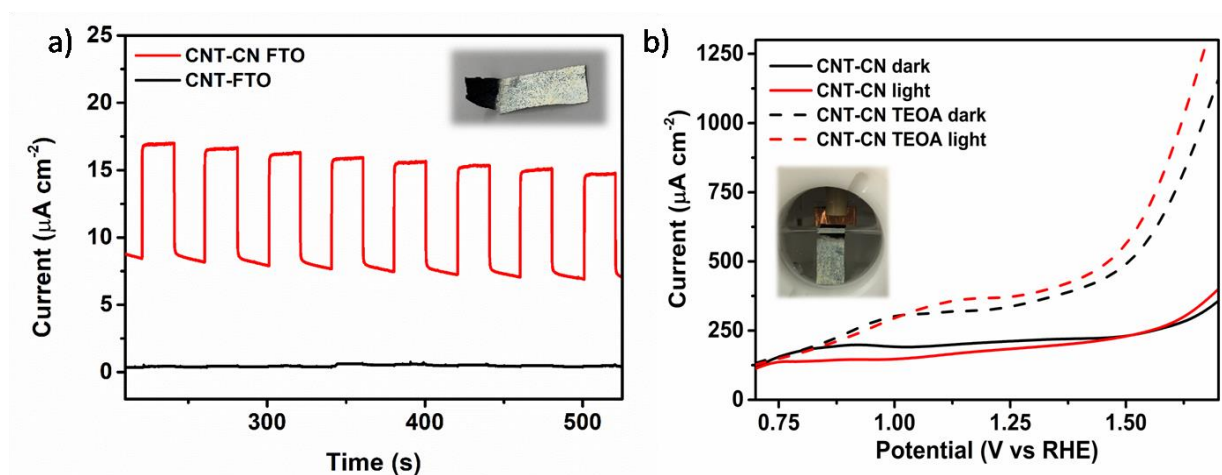

**Figure 6.** Photocurrent measurements of using CNT-CN as photoanodes at 1.23V vs RHE (a) and the subsequent linear sweep voltammetries with and without triethanolamine (b). The figures in the inset show a CNT-CN self-standing electrode and the same electrode within the PEC cell submerged in the electrolyte.

The table of contents entry should be 50–60 words long and should be written in the present tense and impersonal style (i.e., avoid we). The text should be different from the abstract text.

**Controlled Nucleation and Growth of Carbon Nitride Films on CNT Fiber Fabric for Photoelectrochemical Applications**

Neeta Karjule, Moumita Rana, Menny Shalom, Jesús Barrio* and Juan José Vilatela*

Carbon nitride films are grown over CNT through chemical vapor deposition of melamine. CNT acts as a high surface area nucleation center for the formation of oriented carbon nitride which results in enhanced charge transfer and separation properties under visible light irradiation.

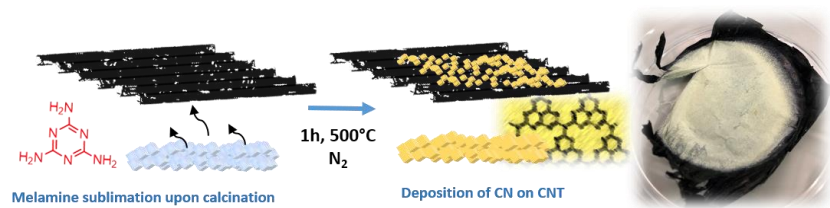

# Supporting Information

Neeta Karjule, Moumita Rana, Menny Shalom, Jesús Barrio* and Juan José Vilatela*

Dr. Neeta Karjule, Prof. Menny Shalom, Dr. Jesús Barrio
Department of Chemistry and Ilse Katz Institute for Nanoscale Science and Technology, Ben-Gurion University of the Negev, Beer-Sheva 8410501, Israel.

Dr. Moumita Rana, Dr. Juan José Vilatela
IMDEA-Materials, Getafe, Madrid E-28906, Spain.
E-mail: juanjose.vilatela@imdea.org

Dr. Jesús Barrio
Faculty of Engineering, Department of Materials, Imperial College London, London, UK.
E-mail: j.barrio-hermida@imperial.ac.uk

Keywords: Carbon nanotube, carbon nitride, photoelectrochemical cells, water splitting, hybrid materials

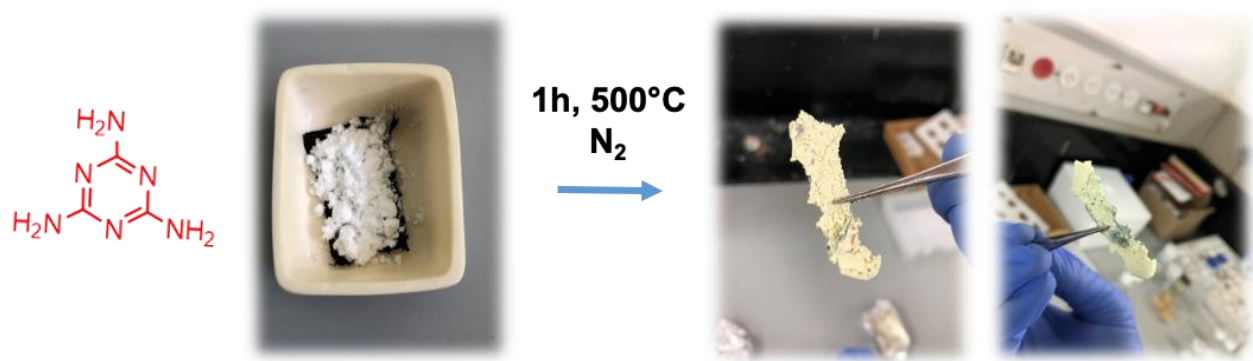

**Scheme S1.** Schematic representation of the preparation of CNT-CN hybrids.

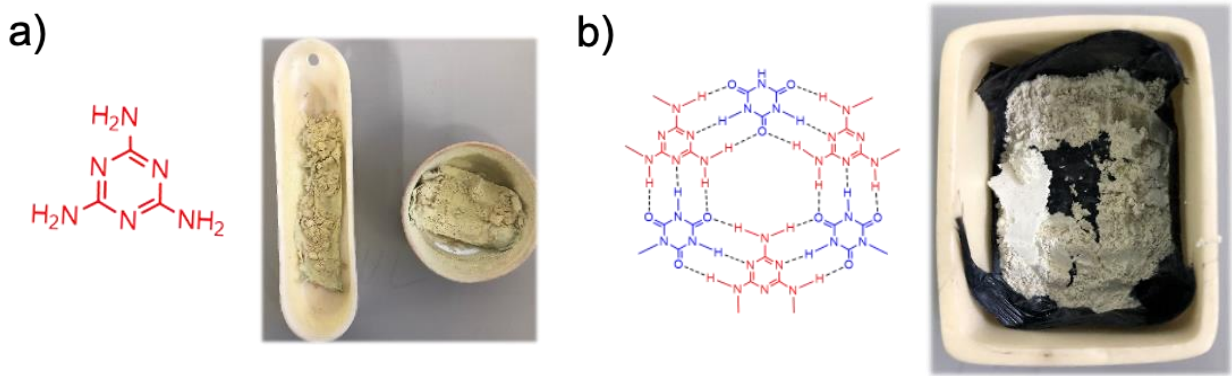

**Figure S1.** CNT-CN materials after thermal condensation when utilizing melamine (a) and the cyanuric acid – melamine complex (b).

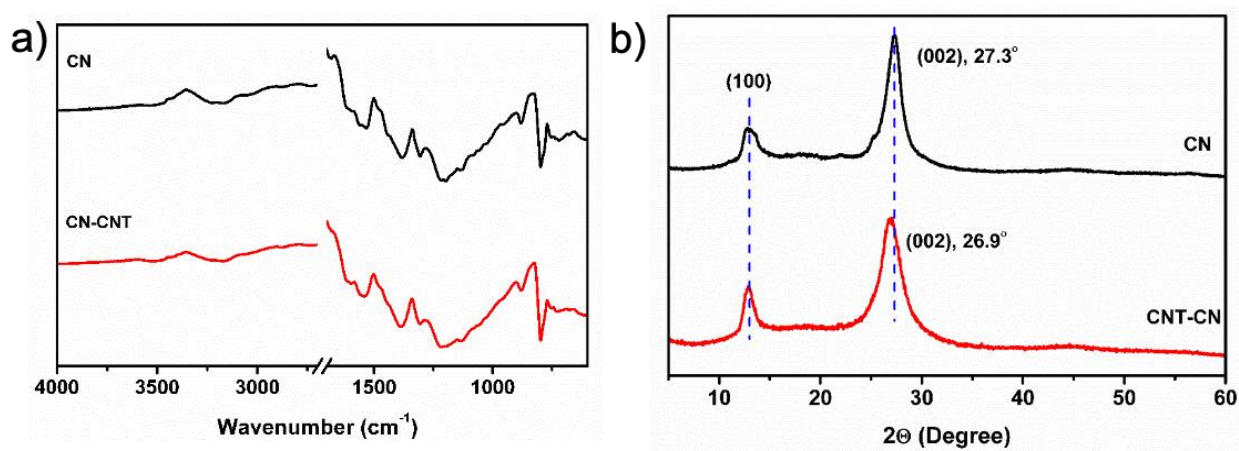

**Figure S2.** FTIR spectra (a) and XRD patterns (b) of CNT-CN materials.

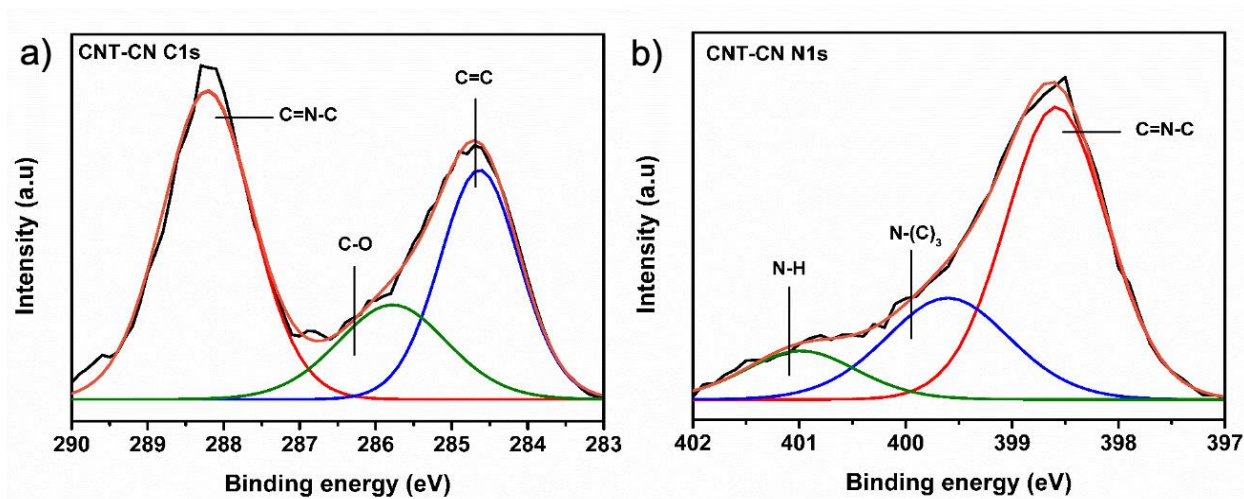

**Figure S3.** XPS spectra for C1s and N1s of CNT-CN. C1s spectra display three different binding energies at 284.8, 285.8 and 288.3 eV which correspond to C-C, C-O, and C-N=C

chemical states. N1s spectra can be deconvoluted in three different binding energies at 398.5, 399.5, and 401.0 eV corresponding to C-N=C, N-C$_3$ and N-H chemical bonds.

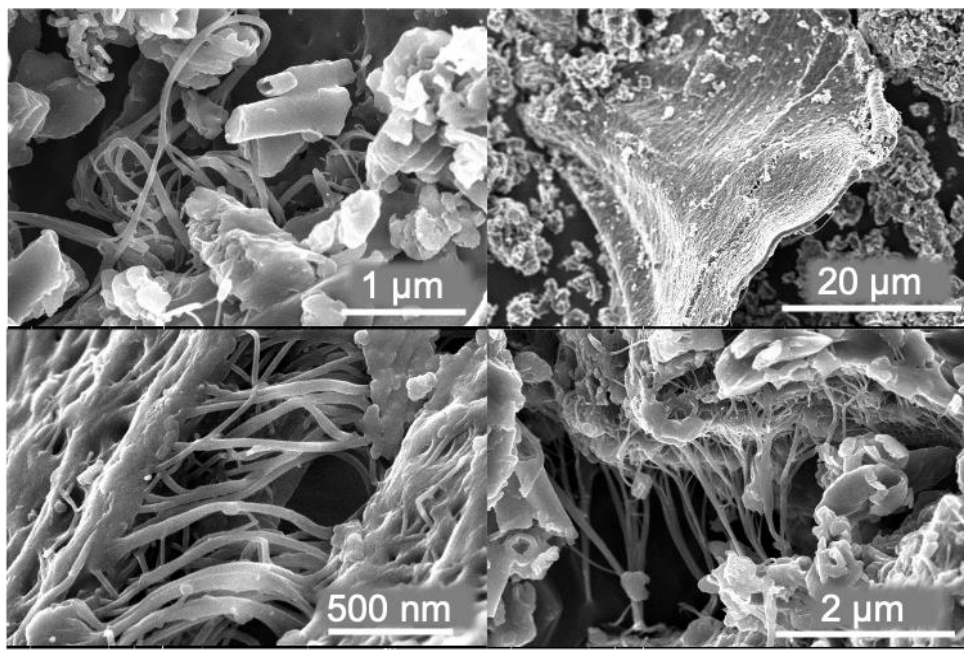

**Figure S4.** SEM images of CNT-CN film grown by direct calcination of melamine on the CNT fiber.

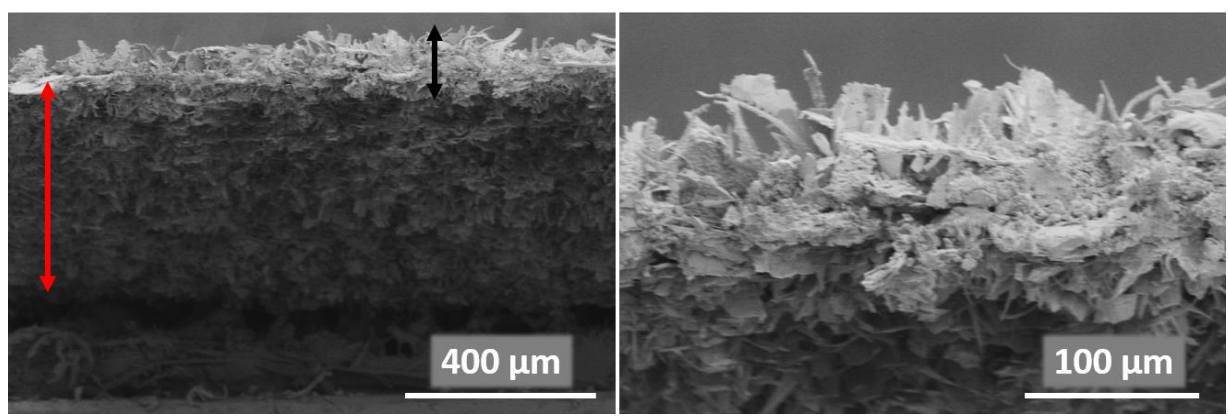

**Figure S5**. Cross-section SEM images of CNT-CN film grown by direct calcination of melamine on CNT fiber fabric. The CN layer is marked with a black double-headed arrow (Thickness: ca. 96 µm).

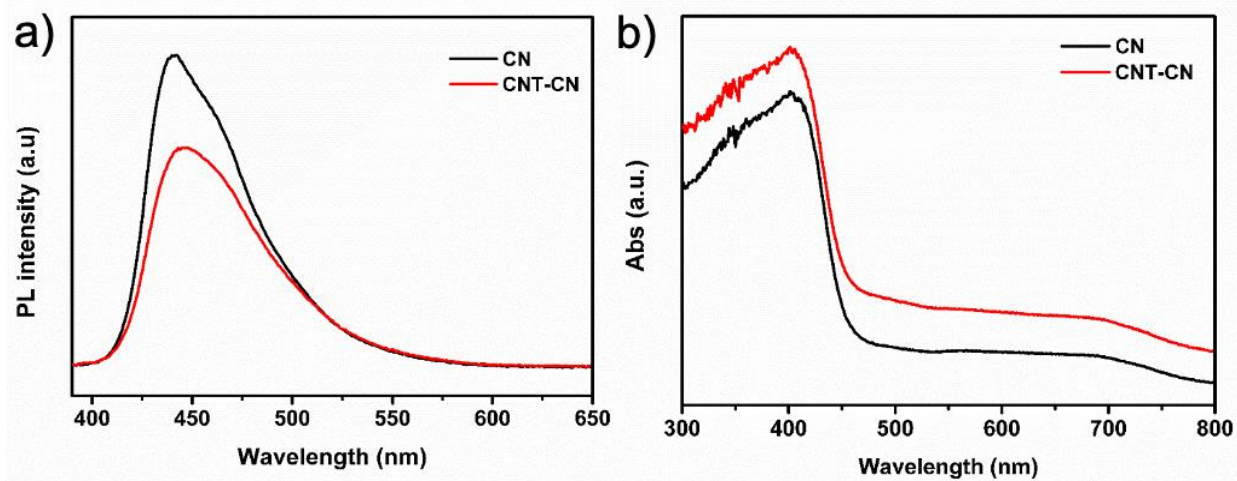

**Figure S6.** Fluorescence (a) and UV-vis (b) spectra of CN and CNT-CN materials.

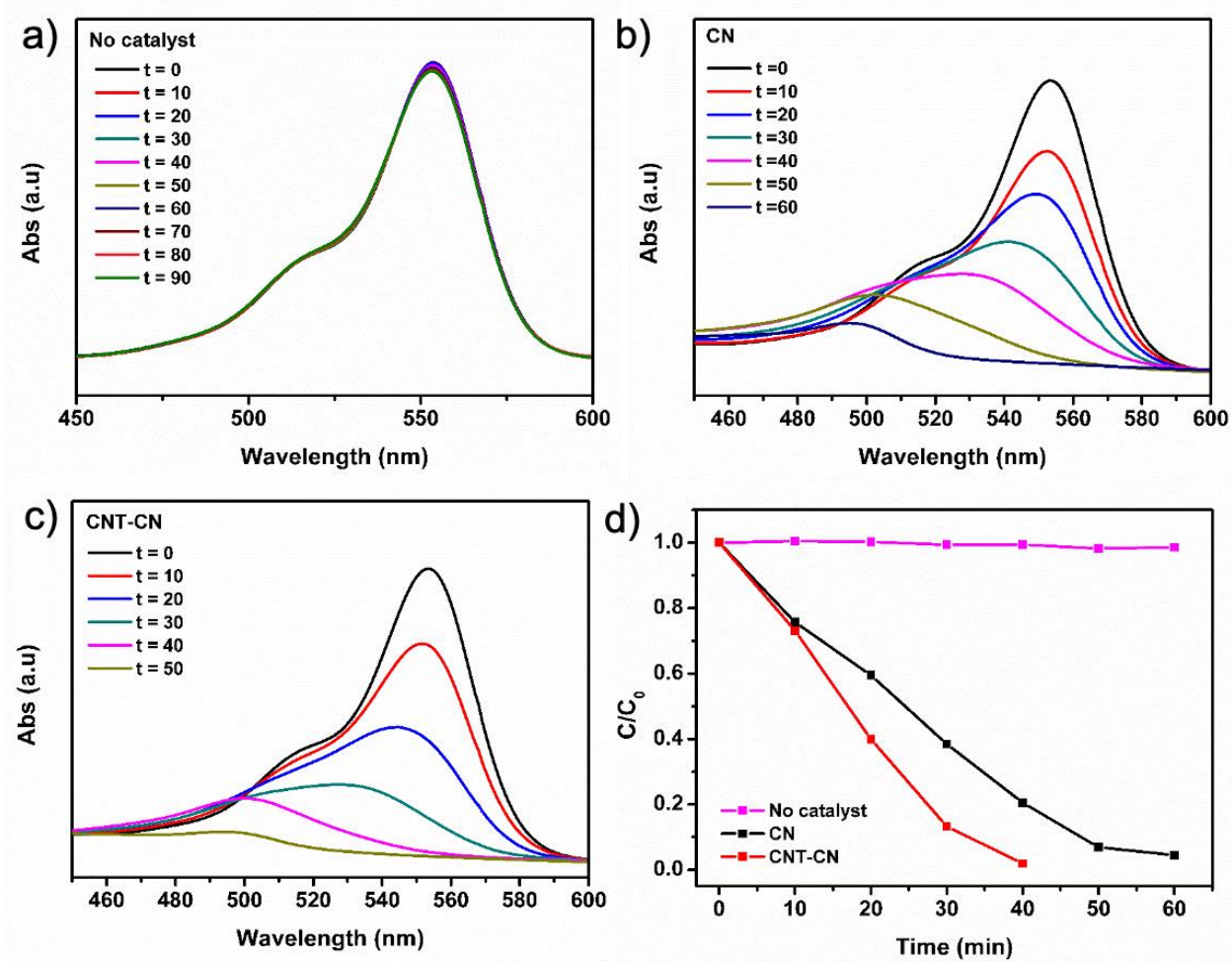

**Figure S7.** UV-vis degradation curves of RhB with no catalyst (a), with CN (b), and with CNT-CN (c). Relative concentrations with different illumination time (d).

We evaluated the rhodamine B degradation performance of the material under illumination with a white LED. The formation of a hybrid material between CNT-CN with improved charge separation properties resulted in a faster degradation (40 min) than when utilizing standard CN (60 min). Furthermore, in the absence of catalyst no degradation could be appreciated.

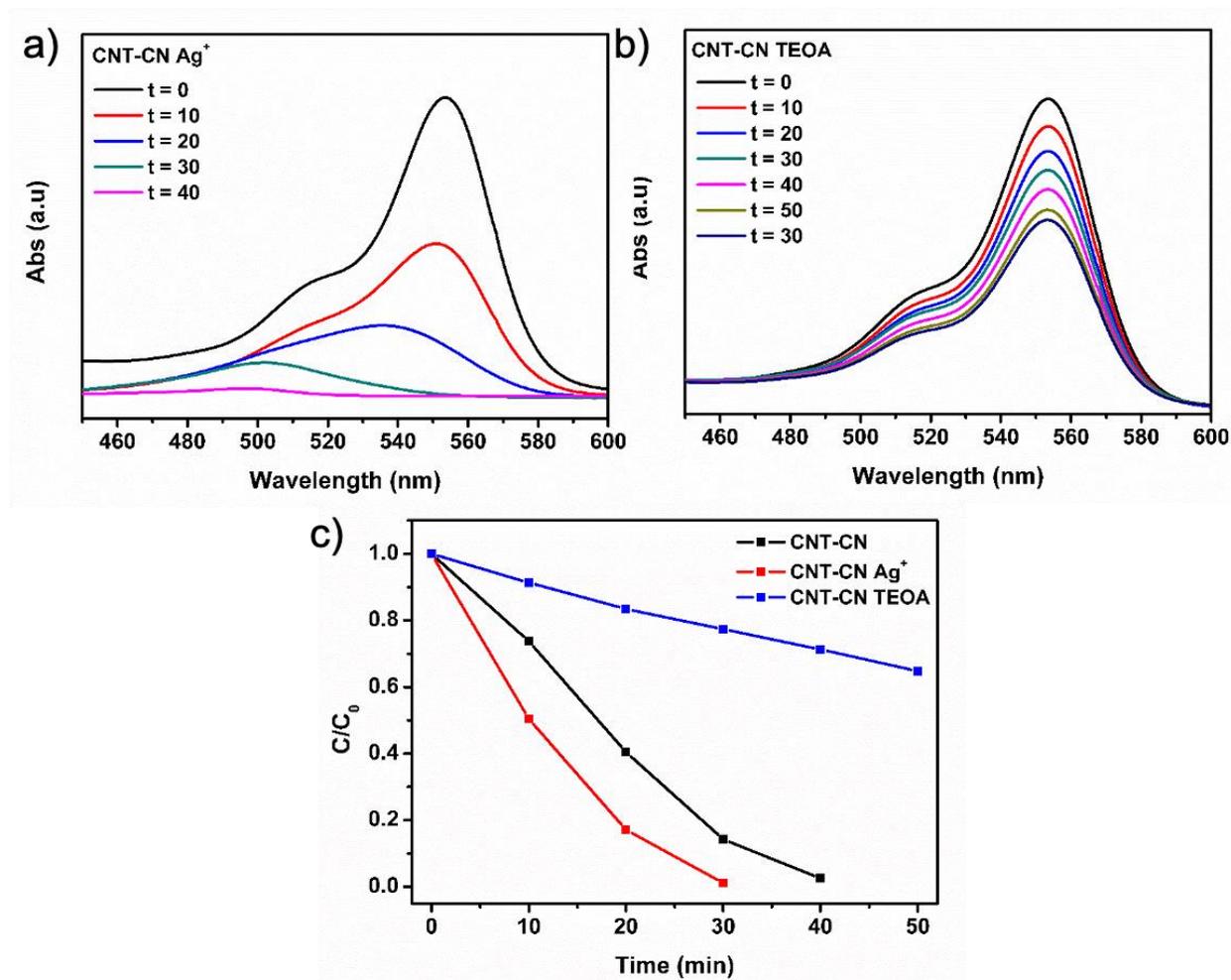

**Figure S8.** UV-vis degradation curves of RhB in the presence of CNT-CN with Ag (a), TEOA (b) and Relative concentrations with different illumination time and scavengers (c).

The photodegradation mechanism was analyzed by performing the photo-catalytic reaction in the presence of scavengers. The addition of $AgNO_3$ to the reaction media accelerates the reaction due to the reduction of $Ag^+$ to $Ag^0$ by photogenerated electrons which improves the lifetime of the photogenerated holes for the formation of hydroxyl radicals or for directly oxidizing rhodamine molecules. This fact was further confirmed by the quench in the degradation efficiency upon addition of triethanolamine (TEOA), which reacts with the holes allowing a higher lifetime to the photogenerated electrons.[1, 2]

**Table S1.** Binding energy values of the different chemical contributions found at XPS

| Level | C1s C-N=C | C1s C-C/C=C | N1s C-N=C | N1s N-C$_3$ |
|---|---|---|---|---|
| Surface | 288.3 eV | 284.8 eV | 398.8 eV | 399.9 eV |
| 250 s, 17.5 nm | 287.7 eV | 284.8 eV | 398.1 eV | 399.1 eV |
| 500 s, 35 nm | 287.9 eV | 284.8 eV | 398.2 eV | 399.3 eV |

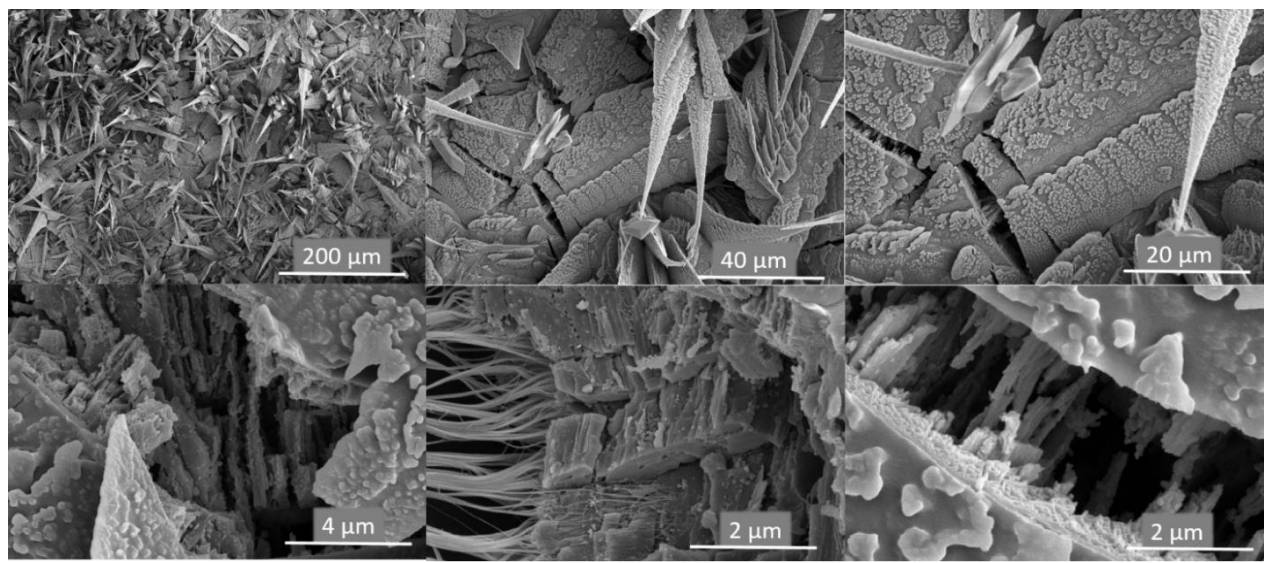

**Figure S9.** SEM images of the CNT-CN films grown by CVD of melamine.

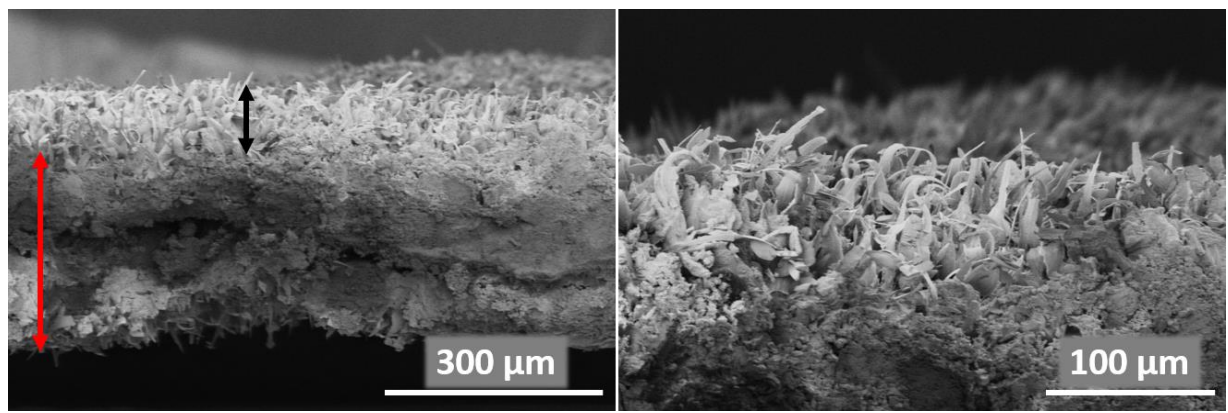

**Figure S10**. Cross-section SEM images of CNT-CN film grown by CVD of melamine. The CN layer is marked with a black double-headed arrow (Thickness: ca. 77 µm).

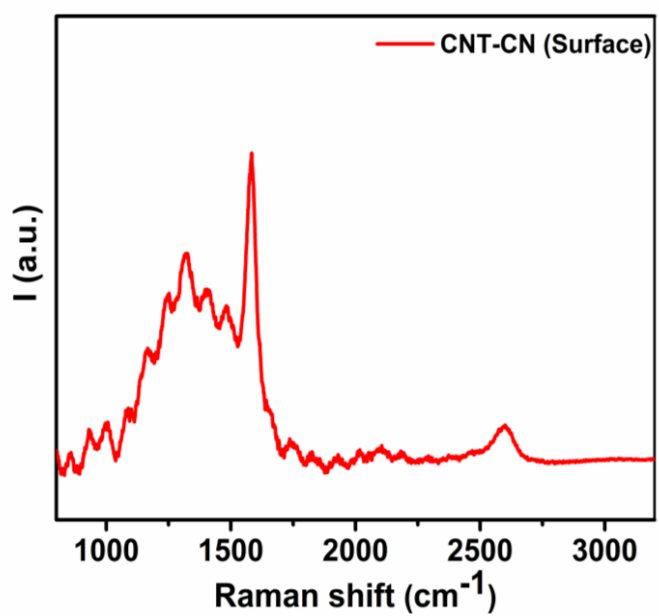

**Figure S11.** Raman spectrum of CNT-CN hybrid.

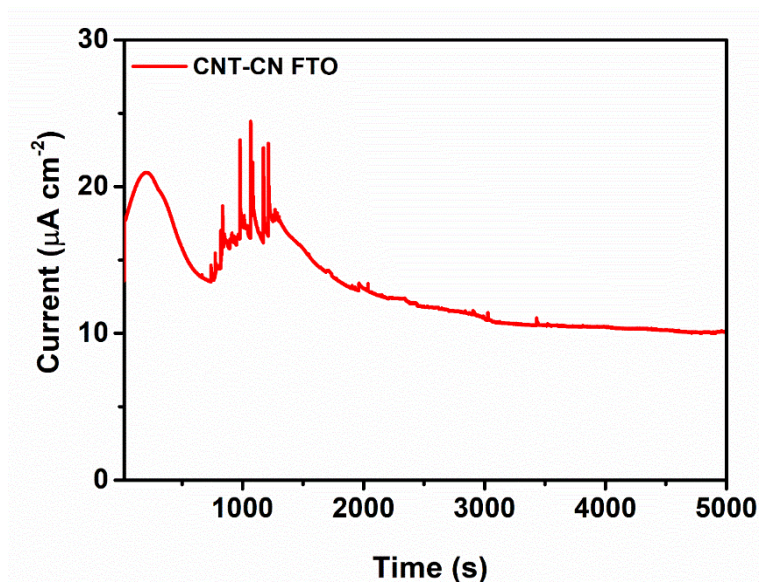

**Figure S12.** Photocurrent stability of CNT-CN at 1.23 V vs RHE in 0.1 M KOH solution.

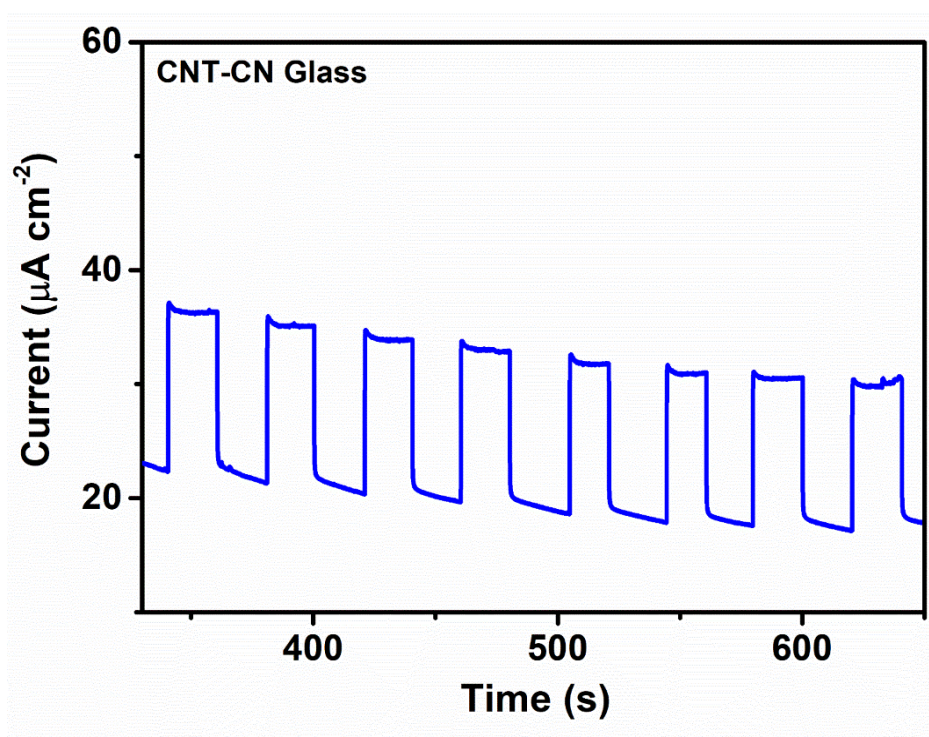

**Figure S13.** Photocurrent measurements of using CNT-CN attached to glass instead of FTO at 1.23 V vs RHE in 0.1 M KOH solution.

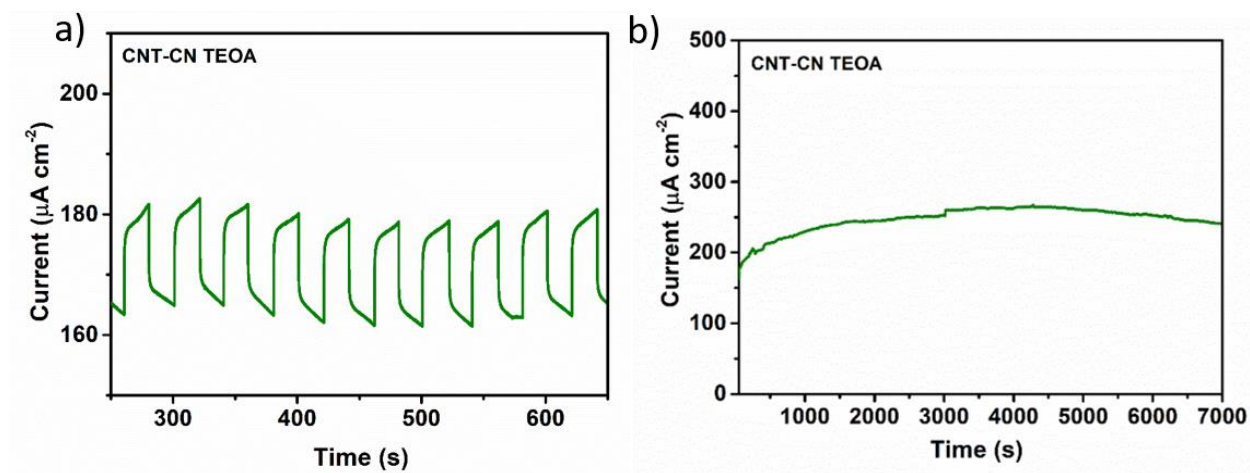

**Figure S14.** Photocurrent (a) and stability (b) photo-electrochemical measurements in the presence of a 0.1 M KOH aqueous solution with 10% (v/v) triethanolamine (TEOA) (at 1.23 V vs RHE).

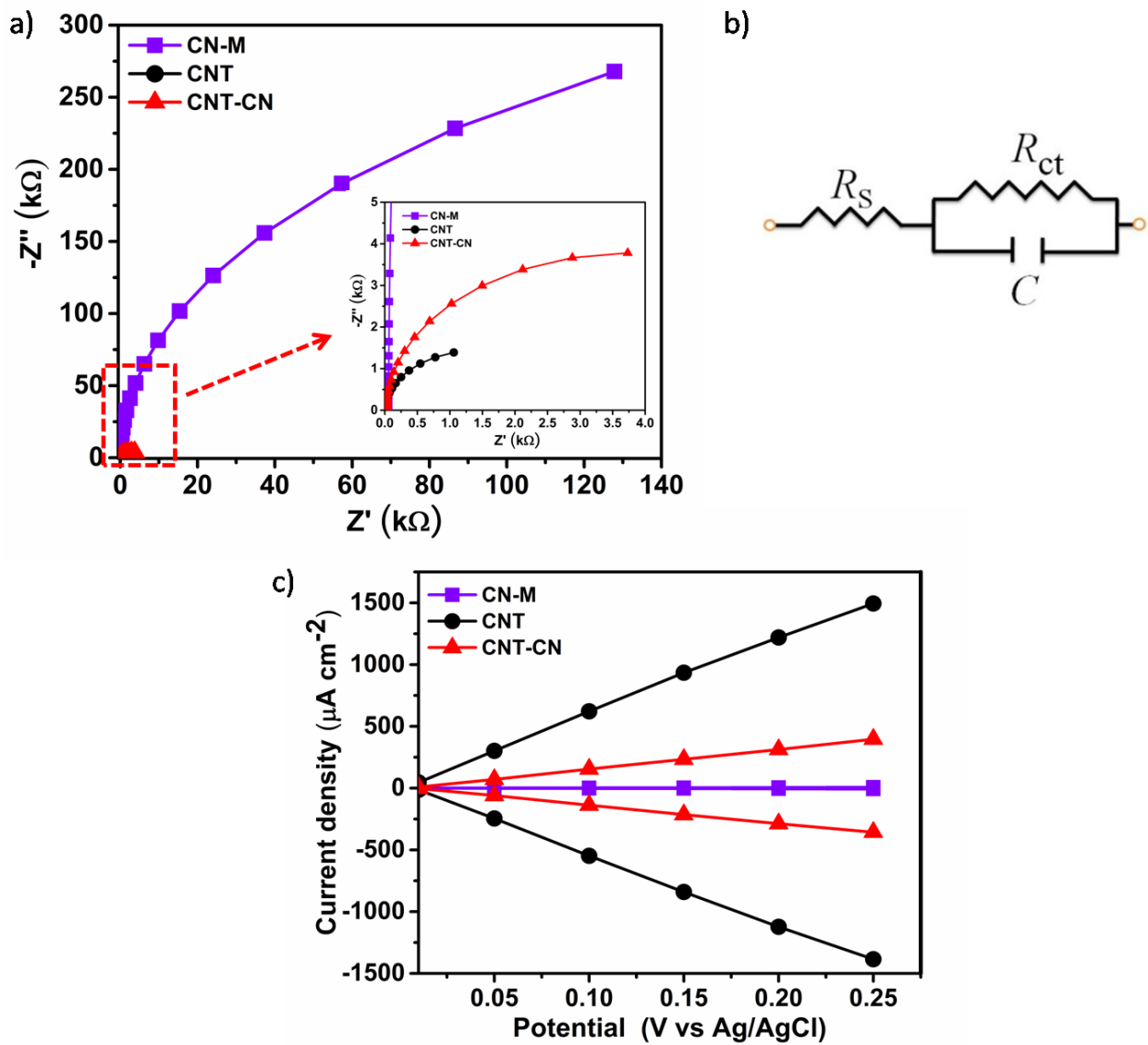

**Figure S15.** Nyquist plots of CN-M (FTO substrate), CNT and CNT-CN at applied potentials of 1.23 V vs RHE in 0.1 M KOH aqueous solution (a), the equivalent circuit for the fitting of Nyquist plots (b), cathodic and anodic charging currents of materials at 0.1 V vs Ag/AgCl as a function of scan rate to calculate the double-layer capacitance ($C_{dl}$) (c).

**Table S2.** Summarized data from EIS and ECSA

| Sample name | $R_{ct}$ (kΩ) data obtained from EIS | $C_{dl}$ (mF cm$^{-2}$) obtained from ECSA |
|---|---|---|
| CN-M | 689.21 | 0.027 |
| CNT | 2.89 | 6.06 |
| CNT-CN | 11.84 | 1.61 |

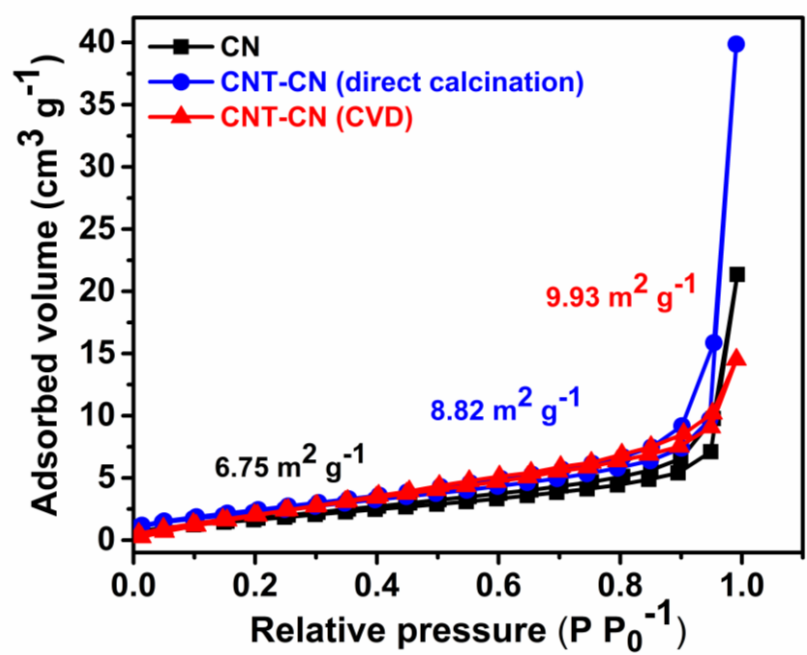

**Figure S16.** N$_2$ adsorption-desorption isotherms for CN (black color curve), CNT-CN (grown by direct calcination of melamine on the CNT, blue color curve) and CNT-CN (grown by CVD of melamine on the CNT, red color curve).

**References**

1　　L. Li, M. Shalom, Y. Zhao, J. Barrio, M. Antonietti, *J. Mater. Chem. A* **2017**, *5*, 18502.
2　　L. Li, Y. Zhao, M. Antonietti, M. Shalom, *Small* **2016**, *12*, 6090.